\documentclass{JHEP3}
\usepackage{amsmath}
\usepackage{amssymb}
\usepackage{graphicx}
\numberwithin{equation}{section}

\def\appendix#1{
  \addtocounter{section}{1}
  \setcounter{equation}{0}
  \renewcommand{\thesection}{\Alph{section}}
 \section*{Appendix \thesection\protect\indent \parbox[t]{11.715cm} {#1}}
  \addcontentsline{toc}{section}{Appendix \thesection\ \ \ #1}
  }

\newcommand{\be}{\begin{equation}}
\newcommand{\ee}{\end{equation}}
\newcommand{\ba}{\begin{aligned}}
\newcommand{\ea}{\end{aligned}}

\def\m1{\left(-1\right)^{F_i}}

\makeatletter
\def\sla@#1#2#3#4#5{{%
  \setbox\z@\hbox{$\m@th#4#5$}%
  \setbox\tw@\hbox{$\m@th#4#1$}%
  \dimen4\wd\ifdim\wd\z@<\wd\tw@\tw@\else\z@\fi
  \dimen@\ht\tw@
  \advance\dimen@-\dp\tw@
  \advance\dimen@-\ht\z@
  \advance\dimen@\dp\z@
  \divide\dimen@\tw@
  \advance\dimen@-#3\ht\tw@
  \advance\dimen@-#3\dp\tw@
  \dimen@ii#2\wd\z@  \raise-\dimen@\hbox to\dimen4{%
    \hss\kern\dimen@ii\box\tw@\kern-\dimen@ii\hss}%
  \llap{\hbox to\dimen4{\hss\box\z@\hss}}}}
\def\slashed#1{%
  \expandafter\ifx\csname sla@\string#1\endcsname\relax
    {\mathpalette{\sla@/00}{#1}}%
  \else
    \csname sla@\string#1\endcsname
  \fi}
\makeatother

\def\ccw{{\hspace{-2.5mm}\unitlength 0.1in
\begin{picture}(1.00,1.00)(8.45,-11.50)
\special{pn 8}%
\special{pa 1120 1120}%
\special{pa 1100 1100}%
\special{pa 1080 1120}%
\special{fp}%
\end{picture}%
\hspace{0.45mm}}}

\newcommand{\beq}{\begin{equation}}
\newcommand{\eeq}{\end{equation}}
\newcommand\beqa{\begin{eqnarray}}
\newcommand\eeqa{\end{eqnarray}}
\newcommand\bea{\begin{array}}
\newcommand\eea{\end{array}}
\newcommand{\COMMENT}[1]{{}}
\newcommand{\nn}{\nonumber}
\newcommand{\neqa}{\nonumber\end{eqnarray}}
\newcommand{\la}{\label}

\newcommand{\G}{{\cal G}}
\newcommand{\F}{{\cal F}}

\renewcommand{\O}{{\cal O}}
\newcommand{\color}[1]{}

\newcommand{\eq}[1]{(\ref{#1})}
\newcommand{\eqs}[2]{(\ref{#1},\ref{#2})}

\renewcommand{\t}{\tilde}

\def\({\left(}
\def\){\right)}
\def\[{\left[}
\def\]{\right]}

\def\<{\langle}
\def\>{\rangle}

\def\d{\partial}

\title{Generalized Scaling Function at Strong Coupling}
\author{Nikolay Gromov\\Service de Physique Th\'eorique, CNRS-URA 2306 C.E.A.-Saclay,%
91191 Gif-sur-Yvette, France; Laboratoire de Physique Th\'eorique
de l'Ecole Normale Sup\'erieure et l'Universit\'e Paris-VI, Paris,
75231, France; St.Petersburg INP, Gatchina, 188 300, St.Petersburg,
Russia; E-mail: \email{nikgromov@gmail.com}}

\abstract{We considered folded spinning string in $AdS_5\times S^5$ background
dual to
the ${\rm Tr}\(D^S\Phi^J\)$ operators of ${\cal N}=4$ SYM
theory. In the limit $S,J\to \infty$ and $\ell=\frac{\pi J}{\sqrt\lambda\log S}$ fixed
we compute the string energy with the 2-loop accuracy in the worldsheet coupling $\sqrt\lambda$
from the asymptotical Bethe ansatz.
In the limit $\ell\to 0$ the result is finite due to the massive cancelations with terms
coming from the conjectured dressing phase. We also managed to compute all leading logarithm terms $\frac{\ell^{2m}\log^n \ell}{\lambda^{n/2}}$ to an arbitrary order in perturbation theory. In particular for $m=1$ we reproduced results of Alday and Maldacena computed from a sigma model.
The method developed in this paper could be used for a systematic expansion in $1/\sqrt\lambda$
 and also at weak coupling.}

\keywords{Duality in Gauge Field Theories}
\preprint{LPTENS 08/31\\SPhT-t08/094}

\begin{document}

\section{Introduction}
In this paper we will consider the $sl(2)$ sub-sector of the AdS/CFT duality
describing the operators of the form ${\rm Tr}\(D^S\Phi^J\)$. This sector is
known to be closed perturbatively to all orders in the gauge coupling. This means that the
operators with $S$ derivatives and $J$ scalar fields mix only with each other under renormalization.
The corresponding mixing matrix in the planar `t Hooft limit is believed to be an integrable
Hamiltonian of an $sl(2)$ spin chain for all values of
the `t Hooft coupling $\lambda$. This assumption drastically
simplifies computation of anomalous dimensions of these operators which could be done by
mean of a Bethe ansatz, based on the S-matrix approach \cite{SH38}. In the $sl(2)$ subsector the asymptotic all-loop Bethe equations read \cite{SH5,be,bes,AFS}
\beqa
\la{AFS}\(\frac{x_k^+}{x_k^-}\)^J=
\prod_{j\neq k}^{S}\(\frac{x_k^+-x_j^-}{x_k^--x_j^+}\)^{-1}
\frac{1-1/(x_k^+ x_j^-)}{1-1/(x_k^- x_j^+)}
\,\sigma^2(u_k,u_j)\,,\la{sl2BAE}
\eeqa
where $x^\pm_k\equiv 2\pi\frac{ u_k+i/2}{\sqrt\lambda}+\sqrt{4\pi^2\(\frac{ u_k+i/2}{\sqrt\lambda}\)^2-1}$ and $\sigma^2$ is the
famous dressing factor \cite{BHL,bes}.
If one solves this equation and finds
set of $u_k$'s the anomalous dimension
is given by
\beq
\gamma(\lambda,S,J)=\frac{\sqrt\lambda}{2\pi}\sum_{j=1}^S\(\frac{i}{x_k^+}-\frac{i}{x_k^-}\)\;.
\la{anom}
\eeq

At the string side of the duality the corresponding state is a folded string living in
$AdS_3\times S^1$ and carrying large angular momenta $S$ and $J$. The energy of the string is
given by $S+J+\gamma(\lambda,S,J)$ via the AdS/CFT duality \cite{Maldacena} and the world-sheet sigma model
coupling is $\lambda^{-1/2}$.

The equations \eq{sl2BAE} are still rather complicated.
To simplify the problem
we will consider the limit introduced in \cite{bgk,ftt} when $J,S\to\infty$
and
\beq
\ell=\frac{\pi J}{\sqrt\lambda\log S}\la{defz}
\eeq
 is fixed.
In this limit the anomalous dimensions scales as $\log S$ \cite{gkp} and one defines the so-called generalized
 scaling function $f(\lambda,\ell)$ by
 \beq
 \Delta-S-J=\gamma=\lambda^{1/2}\frac{f(\lambda,\ell)}{\pi}\log S
 \eeq
or equivalently
\beq
f(\lambda,\ell)=\frac{\gamma(\lambda,\ell,J)\ell}{J}\;.\la{eq:fgamma}
\eeq

We will compute this quantity as an expansion in $1/\sqrt\lambda$ keeping a full functional
dependence on $\ell$.
\beq
f(\lambda,\ell)= f_{cl}(\ell)+\lambda^{-1/2}f_{\rm 1-loop}(\ell)+\lambda^{-1}f_{\rm 2-loop}(\ell)+\dots\;.
\eeq

This object was studied intensively at both strong and weak coupling \cite{ft1,bfst,kot,bgk,be,bern,bes,ben,ftt,ald,krj,rtt,am2,bkk,rt,rt2,SH0}.
The strong coupling expansion is known up to two loops to be
\beqa
f_{\rm cl}(\ell)&=&\sqrt{\ell^2+1}-\ell\la{fcl}\\\
f_{\rm 1-loop}(\ell)&=&\frac{\sqrt{\ell^2+1}-1+2(\ell^2+1)\log\(1+\frac{1}{\ell^2}\)-(\ell^2+2)\log\frac{\sqrt{\ell^2+2}}{\sqrt{\ell^2+1}-1}}{ \sqrt{\ell^2+1}}\la{f1loop}\\
f_{\rm 2-loop}(\ell)&=&-C+\ell^2\(8\log^2 \ell-6\log \ell+q_{02}\)+\O\(\ell^4\)\la{f2loops}\;,
\eeqa
where $C$ is Catalan's constant and $q_{02}$ is some number.
The two-loop term \eq{f2loops} have not been yet computed
for an arbitrary $\ell$. Only a couple of terms in small $\ell$
expansion are known \cite{rt2}.
In this paper we will compute $f_{\rm 2-loop}(\ell)$
 directly from Bethe ansatz \eq{sl2BAE}.
We will see that the result is finite in $\ell\to 0$ limit
only due to massive cancelations with terms coming from the dressing factor.

Our method is similar to \cite{krj}, where the one loop result \eq{f1loop} of \cite{ftt} was confirmed from the Bethe ansatz \eq{sl2BAE}.
We will expand \eq{AFS} first in the classical limit $S\sim J\sim \sqrt\lambda$ \cite{KMMZ}
and then pass to the limit described above.
This order of limits is exactly the same as in
perturbative expansion of the worldsheet sigma model \cite{rt} and we are free from the
potential order-of-limits problem.

It is known that a two-loop computation in Bethe ansatz is qualitatively
more complicated problem then a one-loop computation. At two loops the discreet behavior of the
Bethe roots $u_k$ becomes important \cite{GK}. In this paper we will show how to efficiently
override these difficulties and rewrite \eq{AFS} as a quadratic equation.

Basing on some natural assumptions about the behavior of the strong coupling expansion at small
$\ell$ we managed to compute all the terms of the form $\frac{\ell^{2m}\log^n \ell}{\lambda^{n/2}}$
in $f(\lambda,\ell)$
using just 1-loop result for $f(\lambda,\ell)$.
In a particular case $m=1$ we found a perfect agreement with \cite{am2}.

The paper is organized as follows: in Sec. 2 we expand the Bethe equations in classical limit
and rewrite it as a simple quadratic equation,
in Sec. 3 we focus on the terms coming from the Hernandez-Lopez phase and ``anomaly" contribution,
in Sec. 4 we combine all the contributions together and write down our 2-loop
correction to the scaling function, in Sec. 5 we subtract leading logarithms
at all orders in $1/\sqrt\lambda$, in Sec. 6 we conclude. Appendix A contains
some intermediate computation, in Appendix B we write an expansion in powers of $\ell$
and in Appendix C we give our results in \textsl{Mathematica} syntaxis.

\section{Strong coupling expansion of Bethe equations}
In this section we will expand Bethe equations \eq{AFS}
in the strong coupling limit $\lambda\to\infty$.
We will also keep $S,J\sim \sqrt\lambda$. It is well known that in these
settings the Bethe roots $u_k$ scale like $\sqrt\lambda$~\cite{KMMZ}. It is convenient
to introduce
\beq
x_k\equiv 2\pi \frac{ u_k}{\sqrt\lambda}+\sqrt{4\pi^2\(\frac{ u_k}{\sqrt\lambda}\)^2-1}
\eeq
so that $x_k\sim 1$. Then $x_k^\pm$, which enter  the Bethe equations \eq{sl2BAE}
and the expression for anomalous dimensions \eq{anom},
can be expanded in $1/\sqrt\lambda$
\beq
x^\pm_k=x_k\pm \frac{i\alpha(x_k)}{2}+\frac{\alpha^2(x_k)}{4x_k(x_k^2-1)}\pm\dots
\eeq
where $\alpha(x)=\frac{4\pi}{\sqrt\lambda}\frac{x^2}{x^2-1}$.
It will be very useful to introduce a resolvent
\beq
\G(x)=\frac{1}{J}\sum_j\frac{1}{x-x_j}\;.
\eeq
We will also use $g=\frac{\sqrt\lambda}{4\pi}$ for convenience.

Now we can express in a compact form the expansion of anomalous dimension \eq{anom}. In the notations introduced
above for symmetric distribution of roots it reads
\beq\la{eq:gammaex}
\frac{\gamma(g)}{J}=-\(\left.2 \G+\frac{3\G-3\G'-21\G''-10\G^{(3)}-\G^{(4)}}{384g^2}+\O\(\frac{1}{g^4}\)\)\right|_{x=1}\;.
\eeq
To expand Bethe equations one usually takes $\log$ of both sides first. To fix the branch of the logarithm one should add $2\pi i n_k$ where $n_k$ are some integer numbers called mode numbers \cite{KMMZ}.
The expansion is then straightforward and leads to
\beqa
-\frac{2\pi n_k}{J\alpha(x_k)}&=&\frac{2}{J}\sum_{j\neq k}\frac{1}{x_k-x_j}+\frac{\gamma(g)+J}{J x_k }\nn
+\frac{x_k}{4g^2}\(\frac{x_k^4+4x_k^2+1}{(x_k^2-1)^4}-\frac{\G(1)+3\G'(1)+\G''(1)}{3(x_k^2-1)^2}\)\\ \la{BAEexp}
&+&\frac{{\cal V}_{\rm HL}(x_k)}{J\alpha(x_k)}
+\frac{\pi\rho'(x_k)}{J}\(\coth(\pi\rho)-\frac{1}{\pi\rho}\)
+\O\(\frac{1}{g^3}\)\;.
\eeqa
Let us explain the origin of the different terms.
The first line comes from the Bethe equation with the full dressing phase, except the Hernandez-Lopez phase~\cite{HernandezLopez,GV2} which results in the first term in the second line. The second term in the second line is known under the name of ``anomaly" and comes from the terms in the product with $j-k\sim 1$ \cite{anom}.
In this expansion we noticed that the terms $\G^{(n)}(1/x_k)$
appearing all the way cancels out when the 2-loop dressing phase is taken into account. This cancelation could be a very restrictive condition on the phase and is probably equivalent to the crossing\footnote{this cancelation appears also at higher orders. We thank to P.Vieira for discussing this point.}.

Let us emphasize once more that the 2-loop dressing phase is taken into account, but its contribution is not explicitly seen in \eq{BAEexp}. The resulting equation is much simpler and does not contain $\G^{(n)}(1/x_k)$ terms when we mix expansion of the Bethe equation without dressing phase with 2-loop dressing phase.

In the paper \cite{GV2} a very compact representation of the Hernandez-Lopez phase~\cite{HernandezLopez} was derived which we will use here
\beq\la{eq:GV2}
\frac{{\cal V}_{\rm HL}(x)}{\alpha(x)}=\int_{-1}^{1} \(\frac{1}{x-y}+\frac{1}{x}+\frac{1}{1/y-x}\)
\d_y\(\frac{\G(1/y)+y^2 \G(y)-2 y \G(1)}{g(y^2-1)}\)
\frac{dy}{2\pi}\;,
\eeq
where the integration goes along the upper half of the unit circle $|x|=1$.

The anomaly term (the last term in the second line of \eq{BAEexp}) contains density $\rho$ of the roots $u_k$.
We will use two different densities
\beq\la{eq:rho0}
\rho\equiv\frac{1}{\d u_k/\d k}\;\;,\;\;\varrho\equiv \frac{1}{J\d x_k/\d k}\;,
\eeq
which are trivially related
\beq\la{eq:rho}
\rho(x)=J\alpha(x)\varrho(x)\;\;,\;\;
\varrho(x)=-\frac{\G(x+i0)-\G(x-i0)}{2\pi i}\;,
\eeq
where $\alpha(x)=\frac{x^2}{g(x^2-1)}$.

To proceed one have to specify a particular set of mode numbers $\{n_k\}$.
Different sets of mode numbers will lead to different solutions
of the Bethe ansatz. They correspond to different string motions. The one corresponding to the simplest folded string is \beq
n_k=-1\;\;,\;\;k=1,\dots,S/2\;\; ;\;\;n_k=+1\;\;,\;\;k=S/2+1,\dots,S\;.
\eeq
On the gauge theory side this choice corresponds to the twist $J$ operators (i.e. operators with all Lorentz indices symmetrized and traceless). We see that this set of $n_k$'s respects $x_k\to -x_{S-k}$ symmetry and the resulting distribution of roots should by symmetric with respect to the origin
\beq
\varrho(-x)=\varrho(x)\;\;,\;\;\G(-x)=-\G(x)\;.
\eeq
When $S\to \infty$ the roots are distributed
on two symmetric cuts ${\cal C}=(-b,-a)\cup(a,b)$ with $a\sim 1$
and $b\sim S/\sqrt\lambda$ \cite{krj}. It is important
that the upper limit of the distribution scales like $S/\sqrt\lambda$.
We will also see that the resolvent we introduced scales like $1$
in our limit
\beq
\G(x),\varrho(x)\sim 1\;\;{\rm for}\;\;x\sim 1\;.
\eeq
\subsection{Quadratic equation}
Now we are coming to an important step in our calculation.
We will rewrite \eq{BAEexp} as a quadratic equation.
To convert \eq{BAEexp} into a quadratic equation we are using the standard trick -
we multiply the equation by $\frac{1}{J(x-x_k)}$ and sum over $k$. Using that
\beq\la{eq:gg}
\sum_{k\neq j}\frac{2}{J^2(x-x_k)(x_k-x_j)}=\G^2(x)+\frac{1}{J}\G'(x)\;,
\eeq
where the last term is irrelevant for us since it is suppressed by $1/J$.
We arrive at
\beq\la{eq:quad}
-\frac{c^2(x)}{4}=\G^2(x)+\frac{\gamma+J}{J}\frac{\G(x)}{x}+\frac{\F(x)}{\ell^2}\;.
\eeq
This is our main equation which we will use to compute $f(\lambda,\ell)$.
We introduced $\F(x)=\F_0(x)+\F_{\rm HL}(x)+\F_{\rm An}(x)$ with
\beqa
\F_0(x)&=&\frac{\ell^2}{g^2}\int_{\cal C}\frac{\varrho(y)}{x-y}
\frac{y}{4}\(\frac{y^4+4y^2+1}{(y^2-1)^4}-\frac{\G(1)+3\G'(1)+\G''(1)}{3(y^2-1)^2}\)dy\la{eq:F0}\\
\F_{\rm HL}(x)&=&\frac{\ell^2}{g}\sum_k\frac{1}{J(x-x_k)}\frac{g{\cal V}_{\rm HL}(x_k)}{J\alpha(x_k)}\la{eq:F1}\\ \la{eq:F2}
\F_{\rm An}(x)&=&\ell^2\sum_k\frac{\pi\rho'(x_k)}{J^2(x-x_k)}\(\coth(\pi\rho(x_k))-\frac{1}{\pi\rho(x_k)}\)\;.
\eeqa

So far we did not get a closed algebraic equation on the resolvent $\G$.
We introduced above a new object $c(x)$ defined by
\beq\la{eq:defc}
-c^2(x)=\sum_k\frac{8\pi n_k}{J^2\alpha(x_k)(x-x_k)}=
16\pi\frac{g}{J}\int_a^b\frac{\varrho(y)}{ x-y}\(1-\frac{1}{y^2}\)dy\;,
\eeq
which depends on the resolvent. We see that \eq{eq:quad}
is some complicated nonlinear integral equation. Notice that
$c^2(x)$ is suppressed by $\frac{g}{J}\sim \frac{1}{\log S}$.
The reason why we cannot drop it is that the density $\varrho$ behaves as constant for large $y$ and the
integral gets large contribution of order $\log b\sim\log S$ from large $y$'s (see Appendix A).
Since the main contribution comes from $y\gg 1$ for $x\sim 1$ we can neglect $x$ in the denominator and treat $c(x)$ as a constant! In Appendix A we show that
\beq\la{eq:cz}
c^2=\frac{1}{\ell^2}\;.
\eeq
this is how the quantity $\ell\equiv\frac{\pi J}{\sqrt\lambda\log S}$ enters into our calculation.

We started from a two cut configuration whose resolvent, as is well known, is usually expressed in terms of some elliptic integrals \cite{KMMZ}. However when $S\to \infty$ our two branch points are effectively merging at infinity and we are therefore left with what resembles a single cut solution. This explains why we can still compute the resolvent by solving a  quadractic equation.

\subsection{Resolving quadratic equation}
The equation \eq{eq:quad} with $c(x)= 1/\ell$
becoms a simple quadratic equation.
We can immediately solve it and find $\G(x)$
\beq\la{eq:GF}
\G(x)=\frac{\sqrt{a^2-x^2(1+4\F)}-a}{2\ell x}\;,
\eeq
where we introduced $a$
\beq\la{eq:af}
a\equiv \frac{\gamma(\lambda)   +J}{J}\ell = f(\lambda,\ell)+\ell\;.
\eeq
It is the quantity we are aiming to compute.
$a$ by itself is related to the resolvent and its derivatives
at $x=1$ via \eq{eq:gammaex}. Substituting \eq{eq:GF}
into \eq{eq:gammaex} we will get an algebraical equation on $a$
\beq\la{eq:al}
\ell = \sqrt{a^2-{\cal I}^2}+\frac{8a^4 {\cal I}^4-4a^2 {\cal I}^6+{\cal I}^8}{2^8g^2(a^2-{\cal I}^2)^{7/2}}+\dots\;\;,\;\;{\cal I}\equiv\sqrt{1+4\F(a)}\;,
\eeq
where the dots are standing for some function of $\cal I$ suppressed by $1/g^4$.
The r.h.s. of \eq{eq:al} is some complicated function of $a$. We can try to solve it order by order in $1/g$. Since $\F\sim 1/g$, to the leading order ${\cal I}\simeq 1$ and we have
\beq
a_0=\sqrt{\ell^2+1}\;,
\eeq
which is exactly the classical result \eq{fcl}. To the second order we will get
\beq
a_1=\frac{2\F(1,a_0)}{\sqrt{\ell^2+1}}\la{eq:a1}\;,
\eeq
as we shall see that leads precisely to the correct one-loop result \eq{f1loop} of \cite{ftt,krj}.

For the second order iterations give
\beq\la{eq:aa2}
a_2=-\frac{2\F^2(1,a_0)}{(\ell^2+1)^{3/2}}-
\frac{8\ell^4+12\ell^2+5}{2^8 g^2\ell^6\sqrt{\ell^2+1}}
+\frac{2\d_a\F^2(1,a_0)}{\ell^2+1}\;.
\eeq
In this way we can express $a$ to an arbitrary order
in $\F$. $\F$ by itself is a function of $g$. We will denote
\beq
\F(x)=\delta \F(x)+\tilde \F(x)+\O(1/g^3)\;\;,\;\;\delta \F(x)\sim \frac{1}{g^2}\;\;,\;\;\tilde \F(x)\sim \frac{1}{g}\;.
\eeq
To compute $\F(x)$ via (\ref{eq:F0},\ref{eq:F1},\ref{eq:F2}) we will need to know resolvent $\G(x)$.
The resolvent can be also represented as a series in $\F$ using \eq{eq:GF}
\beq\la{eq:GG}
\G(x)=\t \G(x)+\delta\G(x)+\O\(\F^2\)\;\;,\;\;\t \G(x)\equiv \frac{\sqrt{a^2-x^2}-a}{2\ell x}\;\;,\;\;\delta \G(x)\equiv-\frac{x \F(x)}{\ell\sqrt{a^2-x^2}}\;.
\eeq
Accordingly we also expand the density $\varrho(x)=\t\varrho(x)+\delta\varrho(x)$
\beq\la{eq:rhoexp}
\t\varrho(x)=\frac{\sqrt{x-a}\sqrt{x+a}}{2\pi\ell x}\;\;,\;\;\delta\varrho(x)=x\frac{\F(x+i0)+\F(x-i0)}{2\pi\ell\sqrt{x-a}\sqrt{x+a}}\;.
\eeq

To compute $\tilde \F$ we will use the leading term in the resolvent $\tilde \G(x)$, which does not depend on $\F$. Then we use $\tilde \F$ to compute $\G(x)$ with 1-loop accuracy, which is enough to compute
$\delta \F$. One can continue this iterative procedure to higher orders.

In the Sec. 3 we will compute $\F$ as described above.
A reader could skip the next section and continue from Sec. 4 where
the results are summarized and are used to compute $f(\lambda,\ell)$.

\section{Computation of $\F$}
\subsection{Hernandez-Lopez phase contribution}
In this section we will calculate the contribution of the Hernandez-Lopez phase \eq{eq:F1}.
Using \eq{eq:GV2} we can write
\beqa\la{eqF1}
&&\F_{\rm HL}(x)=\frac{\ell^2}{g}\sum_k\frac{1}{J(x-x_k)}\frac{g{\cal V}_{\rm HL}(x_k)}{J\alpha(x_k)}=\\
&&\nn\frac{\ell^2}{g}
\int_{-1}^{1}\!
\(\frac{\G(x)}{x}\!-\!\frac{\G(x)-\G(1/y)}{x-1/y}\!+\!\frac{\G(x)-\G(y)}{x-y}\)
\!\d_y\!\(\frac{\G(1/y)+y^2 \G(y)-2 y \G(1)}{y^2-1}\)\!
\frac{dy}{2\pi}\;,
\eeqa
where the path of integration goes along upper half of the unit circle $|x|=1$.

To calculate $\F_{\rm HL}(x)$ to the leading order in $g$ one just replaces $\G(x)$ by $\t\G(x)$
from \eq{eq:GG} which we denote by $\t \F_{\rm HL}(x)$. A straightforward integration leads to\footnote{One can copy \eq{eq:tF1} directly to \textsl{Mathematica} from Appendix C, Tab.\ref{tab:1}.
}
\beqa
&&\nn\t \F_{\rm HL}(x)=-\frac{(a^2-1)}{4\pi g(x^2-1)^2}\(2\frac{x^2-1}{a^2-1}+
4\sqrt{\frac{a^2-x^2}{a^2-1}}\log\frac{a^2}{a^2-1}
+\frac{2a^2-x^2-1}{a^2-1}\log\frac{a^4}{a^4-1}\right.\\
&&+\la{eq:tF1}
2\sqrt{\frac{x^2-a^2}{a^2x^2-1}}\frac{a^2x^2+a^2-2}{a^2-1}\[
\tan^{-1}\(\frac{\sqrt{1-a^2x^2}}{\sqrt{a^2-1}}\)-
\tan^{-1}\(\frac{\sqrt{1-a^2x^2}}{\sqrt{a^2-x^2}}\)
\]\\
&&-\left.\nn
\frac{2}{x}\[\frac{a^2+a^2 x^2-2x^2}{a^2-1}+2(x^2+1)\sqrt{\frac{a^2-x^2}{a^2-1}}\]
\[\tanh^{-1}(x)-\tanh^{-1}\(\frac{x\sqrt{a^2-1}}{\sqrt{a^2-x^2}}\)\]\!\)\;.
\eeqa
\subsubsection{Subleading order}
To the next order we need $\F_{\rm HL}(x)$ only for $x=1$, according to \eq{eq:a2}. In this case we can simplify \eq{eqF1} further.
\beq
\F_{\rm HL}(1)=\frac{\ell^2}{2\pi g}\int_{-1}^1 \d_y \(\frac{y^2\G(y)+\G(1/y)-2y\G(1)}{y^2-1}\)\(\frac{y\G(y)+y\G(1/y)-2\G(1)}{y^2-1}\) dy\;.
\eeq
Substituting $\G(x)=\t\G(x)+\delta\G(x)$ and taking the linear in $\delta \G$ term, after integration
by parts we find
\beqa
\delta \F_{\rm HL}(1)&=&\frac{\ell^2}{g}\int_{-1}^1 \(\frac{2C(y) \delta\G(1)}{y}+C(1/y)\delta\G(1/y)-C(y)\delta\G(y)\)\frac{dy}{4\pi y}\;,
\eeqa
where
\beq
C(y)=\frac{y^2}{\ell(y^2-1)^2}\(2\sqrt{a^2-1}-\sqrt{a^2-y^2}-\frac{a^2-1}{\sqrt{a^2-1/y^2}}\)\;.
\eeq
Changing coordinates $y \to 1/y$ in the second term and deforming the contour to the real axe we will get the following
very simple expression
\beq
\delta \F_{\rm HL}(1,a)=\frac{\ell^2}{\pi g}{\rm Re} \[\int_{0}^1 \(\frac{ \delta\G(1)}{y}-\delta\G(y)\)\frac{C(y)dy}{y}\]\;.
\la{eq:F1res}
\eeq
We need only $\delta \G(x)$ to be computed. This will be achieved in the
next section.

\subsection{Anomalous contribution}\la{sec:anom}
The equation \eq{eq:F2} should be understood in the following sense.
We first expand formally \eq{eq:F2} in powers of $1/g$ and then
perform summation over $k$\footnote{This simple prescription was worked
out based on the Airy function behavior of the resolvent close to the branch points~\cite{GK}
in collaboration with Andrzej Jarosz. This prescription was derived for
$sl(2)$ Heisenberg spin chain only. Here we are assuming that it is still valid
for the all-loop $sl(2)$ Bethe ansatz. That could be done since the near branch point behavior
is very universal.
}. To sum over $k$ one can use that
the expression which we have to sum has no poles on the cut $\cal C$
and we can simply multiply it by the resolvent and integrate around
the contour encircling only the singularities of the resolvent $\G$
\beq\la{eq:F22}
\F_{\rm An}(x)= \frac{\ell^2}{J}\ccw\oint_{\cal C}
\(\frac{\pi\d_y\t\rho
\[\coth\(\pi\t\rho\)-1/\pi\t\rho\]}{x-y}+
\frac{\d_y\(\[\coth(\pi\t\rho)-1/\pi\t\rho\]\pi\delta\rho\)}{x-y}\)\G(y)\frac{dy}{2\pi i}\;.
\eeq

At the next stage we have also to expand $\G$.
Each term in the expansion in $1/g$ will have a branch cut instead of a
collection of poles at positions of the Bethe roots.
The sub-leading $1/g$ term in the expansion should behave as
$-\frac{1}{4J(x-a)}$ close to the branch points as we shall see (see also \cite{GK}).
This term is $g/J$ suppressed and thus is missing in the above analysis which was done
to the leading order in $g/J$. To see this near branch point behavior we have to go back
to the equation \eq{BAEexp} and rewrite it in the continuous limit as
\cite{KMMZ,GK}
\beq
\frac{2\pi n}{J\alpha(x)}=-2\slashed \G-\frac{\gamma(g)+J}{Jx}-\frac{{\cal V}_{\rm HL}(x_k)}{J\alpha(x_k)}-\frac{\pi\rho'(x)\coth(\pi\rho)}{J}+\O(1/g^2)\;,
\eeq
where
\beq
\slashed\G(x) \equiv \frac{\G(x+i0)+\G(x-i0)}{2}\;.
\eeq
Close to a branch point density goes to zero as a square root $\rho\sim \sqrt{x-a}$. The last term becomes singular and we have
\beq
\slashed \G(x)\simeq-\frac{\pi \rho'\coth(\pi\rho)}{2J}\simeq-\frac{1}{4J(x-a)}
\eeq
which proves our claim. For more details about behavior of resolvents
near branch points we refer to \cite{GK}.

Although this singularity in $\G$ is suppressed by $g/J$
it will lead to a finite contribution which we call ``boundary term".

\subsubsection{Boundary term}
We replace
$\G$ in \eq{eq:F22} by $-\frac{1}{4(y-a)}-\frac{1}{4(y+a)}$.
The contour of integration now contains only 2 poles inside
and we just have to evaluate the expression in the brackets an $y=\pm a$.
Consider first the contribution from $y=a$.
\beq
-\ell^2\frac{\pi\d_y\rho[\coth(\pi\rho)-1/\pi\rho]}{4J^2(x-a)}
\simeq-\frac{\ell^2}{4J^2(x-a)}\d_x\frac{\pi^2\rho^2}{6}
=-\frac{a^3}{48g^2(a^2-1)^2(x-a)}\;.
\eeq
Taking into account a similar contribution from $x=-a$ we will get
\beq
\F_{\rm An}^{\rm boundary}(x)=\frac{a^4}{24 g^2(a^2-1)^2(a^2-x^2)}\;.
\eeq
We see that all factors of $J$ cancel and we get a finite contribution.
For
$x=1$ and $a=a_0=\sqrt{\ell^2+1}$ we get
\beq
\F_{\rm An}^{\rm boundary}(1,a_0)=\frac{(\ell^2+1)^2}{24g^2\ell^6}\;.\la{eq:F2res0}
\eeq
We see that this term is very singular in the limit $\ell\to 0$.
However then we add all pieces together the full result is completely finite
as we shall see.

\subsubsection{Bulk contribution}
In this section we will drop poles of the resolvent at
the branch points. This implies that we can pass to the integration along the cut with density $\varrho(x)$
\beq
\la{eq:Fblk}\F_{\rm An}^{\rm bulk}(x)\equiv \ell^2\int_{\cal C}
\(\frac{\pi\d_y\t\rho
\[\coth\(\pi\t\rho\)-1/\pi\t\rho\]}{x-y}+
\frac{\d_y\(\[\coth(\pi\t\rho)-1/\pi\t\rho\]\pi\delta\rho\)}{x-y}\)\frac{\varrho(y)dy}{J}\;.
\eeq
Where we use notations introduced above
\beq
\t\rho(x)=J\alpha(x)\t\varrho(x)\;\;,\;\;
\delta\rho(x)=J\alpha(x)\delta\varrho(x)\;\;,\;\;
\alpha(x)=\frac{x^2}{g(x^2-1)}\;.\la{eq:rhorem}
\eeq
In \eq{eq:Fblk} there are contributions of both $1/g$ and $1/g^2$ orders. We split $\F_{\rm An}^{\rm bulk}(x)$ further into $\t \F_{\rm An}(x)$ and $\delta \F_{\rm An}^{\rm bulk}(x)$ as defined below.
\beq
\t \F_{\rm An}(x)\equiv \ell^2\int_{\cal C}
\frac{\pi\t\varrho\d_y\t\rho
\[\coth\(\pi\t\rho\)-1/\pi\t\rho\]}{x-y}
\frac{dy}{J}=
\ell^2\int_{\cal C}
\frac{\pi\t\varrho\d_y\t\rho}{x-y}
\frac{dy}{J}\;,\la{eq:F2b}
\eeq
where in the last equality we use that from \eq{eq:rhorem} $\t\rho(y)\sim J/g\gg 1$
which allowed us to replace $[\coth(\pi\t\rho)-1/\pi\t\rho]$ by $1$ in the second equality.
Using \eq{eq:rhoexp} one can easily evaluate the integral \eq{eq:F2b}
to get\footnote{One can copy \eq{eq:tF2} directly to \textsl{Mathematica} from Appendix C, Tab.\ref{tab:1}.
}
\beq\la{eq:tF2}
\t \F_{\rm An}(x)=\frac{x\log\frac{a-1}{a+1}\(1+x^2-2a^2\)+2a x (x^2-1)+
\log\frac{a-x}{a+x}\(a^2 x^2+a^2-2x^2\)}{4\pi g x(x^2-1)^2}\;.
\eeq
\subsubsection{Second order}
The last contribution of $1/g^2$ order into $\F_{\rm An}(x)$ reads
\beqa
\nn\delta \F_{\rm An}^{\rm bulk}(x)&\equiv& \ell^2\int_{\cal C}
\(\frac{\d_y\t\rho
\[\coth(\pi\t\rho)-1/\pi\t\rho\]\pi\delta\varrho}{x-y}+
\frac{\t\varrho\d_y\!\(\[\coth(\pi\t\rho)-1/\pi\t\rho\]\pi\delta\rho\)}{x-y}\)\frac{dy}{J}\\
&=&\ell^2\int_{\cal C}\frac{\d_y\(\pi\delta\rho\t\rho\[\coth(\pi\t\rho)-1/\pi\t\rho\]\)}{x-y}\frac{dy}{J^2\alpha(y)}\\ \nn
&=&-\int_{\cal C}\d_y\(\frac{1}{\alpha(y)(x-y)}\)\frac{\pi\ell^2\delta\rho\t\rho dy}{J^2}\;.
\eeqa
To evaluate this integral we need $\delta\rho$ which can be expressed in terms of $\F$ \eq{eq:rhoexp}.
We have
\beq
\frac{\pi\ell^2\delta\rho\t\rho}{J^2}=y^4\frac{\t\F(y+i0)+\t\F(y-i0)}{4\pi g^2(y^2-1)^2}\;.
\eeq
Setting $x=1$ we will get the following simple result
\beqa\la{eq:F2res}
\delta \F_{\rm An}^{\rm bulk}(1,a)=-\frac{1}{g}\int_{\cal C}\frac{\t \F(y+i0)+\t \F(y-i0)}{4\pi}\frac{y^2 dy}{(y^2-1)^2}\;,
\eeqa
where $\t \F=\t \F_{\rm HL}+\t \F_{\rm An}$. Using \eq{eq:tF1} and \eq{eq:tF2} one can see that
\beqa
\nn&&\t \F(x+i0)+\t \F(x-i0)=\frac{1}{\pi g(x^2-1)^2}\((a-1)(x^2-1)+
\frac{1+x^2-2a^2}{2}\log\frac{(a-1)a^4}{(a+1)(a^4-1)}
\right.\\ \nn
&&+\frac{a^2x^2+a^2-2x^2}{2x}\log\frac{(x+1)(x-a)}{(x-1)(x+a)}
+(2-a^2-a^2x^2)\sqrt{\frac{x^2-a^2}{a^2x^2-1}}\arctan\sqrt{\frac{x^2-a^2}{a^2x^2-1}}\\
&&\left.+2\frac{x^2+1}{x}\sqrt{(a^2-1)(x^2-a^2)}\arctan\sqrt{\frac{x^2-a^2}{x^2(a^2-1)}}\)\la{eq:stF}\;.
\eeqa
The integral \eq{eq:F2res} can be computed numerically for an arbitrary value of $a$\footnote{In Appendix C in Tab.\ref{tab:2} we give a \textsl{Mathematica} code which computes this integral numerically.} or expanded in powers of $\ell$.
The result of this expansion is give in eq.\eq{eq:F2exp}.
\subsection{Computation of $\F_0$}
The only piece left to compute is $\F_0$ \eq{eq:F0}. Since it is already suppressed by $1/g^2$ this
contribution is especially simple to compute. We immediately evaluate integration using
\eq{eq:rhoexp}
\beq
\F_0(1)=-\frac{24a^4+32a^2-7}{2^93(a^2-1)^3g^2}= -\frac{24\ell^4+80\ell^2+49}{2^93 g^2\ell^6}
\la{eq:F0res}
\eeq

\section{Scaling function at one and two loops}\la{sec:2loops}
Using expressions for $a_1$ and $a_2$ in terms of $\F$ \eqs{eq:a1}{eq:a1}
and results of the previous section, where $\F$ was computed up to $1/g^2$
order we will compute the generalized scaling function $f(g,\ell)$
with the two-loop accuracy in this section.
\subsection{One-loop order}
Having $\t \F=\t \F_{\rm HL}+\t \F_{\rm An}$ computed we can immediately compute
the one-loop energy density using \eq{eq:a1}
\beq
f_{\rm 1-loop}(\ell)=\left.8\pi g\frac{\t \F_{\rm HL}(1)+\t \F_{\rm An}(1)}{\sqrt{\ell^2+1}}\right|_{a=\sqrt{\ell^2+1}}\;.
\eeq
From  \eqs{eq:tF1}{eq:tF2} we have for $x=1$
\beq\la{eq:tFof1}
\t \F_{\rm HL}+\t \F_{\rm An}=\frac{2(a-1)+4 a^2\log \frac{a^2}{a+1}
+\log \frac{(a-1)^2}{a^2+1}-a^2\log (a-1)^2(a^2+1)}{4\sqrt\lambda}\;,
\eeq
and we precisely reproduce \eq{f1loop} by setting $a=a_0=\sqrt{\ell^2+1}$!
\subsection{Two-loop order}
\FIGURE[ht]{
    \centering
        \resizebox{90mm}{!}{\includegraphics{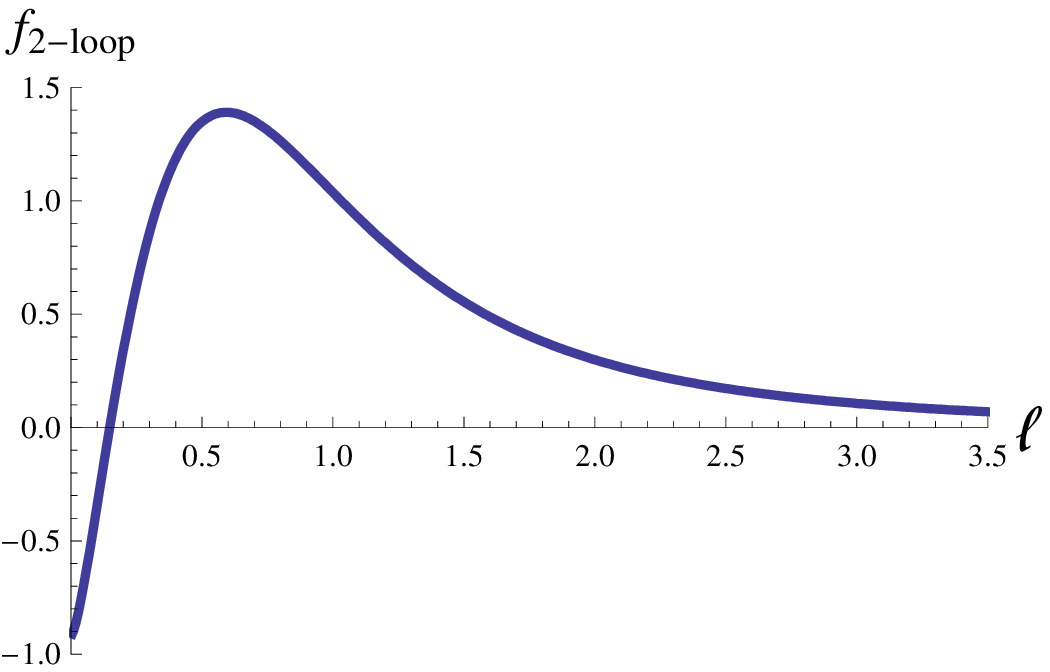}}
        \caption{Two-loop correction to the generalized cusp anomalous dimension as a function of $\ell = \frac{\pi J}{\sqrt\lambda\log S}$. It interpolates between minus Catalan's constant $-C\simeq -0.916$ at small $\ell$'s and $0$ at large $\ell$'s.\la{figH0}}
}
Now we can write down our 2-loop result.
From \eqs{eq:af}{eq:a1} and \eq{eq:aa2} we have
\beq\la{eq:a2}
f_{\rm 2-loop}=\frac{16\pi^2}{\sqrt{\ell^2+1}}\(\frac{2g^2\d_a\t\F^2(a_0)}{\sqrt{\ell^2+1}}-\frac{2g^2\t\F^2(a_0)}{\ell^2+1}+2g^2\delta\F-
\(\frac{5}{256\ell^6}+\frac{3}{64\ell^4}+\frac{1}{32\ell^2}\)
\)\;,
\eeq
where $a_0=\sqrt{\ell^2+1}$ and
\beqa
\t \F&=&\t \F_{\rm HL}+\t \F_{\rm An}\\
\delta \F&=&\F_0+\delta\F_{\rm HL}+\delta\F_{\rm An}^{\rm bulk}+\delta\F_{\rm An}^{\rm boundary}\;.\la{eq:resFF}
\eeqa
The quantities in the r.h.s. of the first line are given by \eqs{eq:tF1}{eq:tF2} and
of the second line by (\ref{eq:F0res},\ref{eq:F1res},\ref{eq:F2res0},\ref{eq:F2res}).
$\delta \F_{\rm HL}$ and $\delta \F_{\rm HL}^{\rm boundary}$ could be represented explicitly as single integrals.
To evaluate them numerically one can use the \textsl{Mathematica} code form Tab.\ref{tab:3} of Appendix C.
In Appendix B we give an expansion of these integrals in power series in $\ell$ up to $\ell^6$ order.

Let us see that the result \eq{eq:a2} is finite in the small $\ell$ limit. This will be already
a very nontrivial test of our calculation because a priory the r.h.s. is divergent as $1/\ell^6$.
For the expansion in $\ell$ we have
\beqa
\frac{2g^2\d_a\t \F^2}{\sqrt{\ell^2+1}}&\simeq&\frac{\log8\log\ell}{4\pi^2}+\frac{\log^28-\log 8}{16\pi^2}\\
-\frac{2g^2\t \F^2}{\ell^2+1}&\simeq&-\frac{\log^28}{32\pi^2}\\
2g^2 \F_0&=&
-\frac{49}{768\ell^6}-\frac{5}{48\ell^4}-\frac{1}{32\ell^2}\\
2g^2 \delta\F^{\rm boundary}_2&=&
\frac{1}{12\ell^6}+\frac{1}{6\ell^4}+\frac{1}{12\ell^2}
\eeqa
Using expansion from Appendix B we have
\beqa
&&2g^2\delta \F_{\rm HL}\simeq
 \frac{1}{\ell^4} \left(-\frac{1}{4 \pi ^2}+\frac{1}{24 \pi }+\frac{\log 8}{24 \pi ^2}\right)
+\frac{1}{\ell^2}\left(-\frac{1}{48}-\frac{1}{2 \pi ^2}+\frac{1}{8 \pi }\right)\\ \nn
&&+\left(-\frac{\log\ell}{32 \pi }-\frac{5 \log 8\log\ell}{16 \pi ^2}-\frac{\log^28}{96 \pi ^2}-\frac{\log 8}{64 \pi ^2}+\frac{\log 8}{64 \pi }+\frac{9}{128 \pi }-\frac{C}{8 \pi ^2}-\frac{1}{16 \pi ^2}-\frac{1}{64}\right)\\
&&2g^2\delta \F^{\rm bulk}_{\rm An}\simeq
\frac{1}{\ell^4} \left(-\frac{1}{64}+\frac{1}{4 \pi ^2}-\frac{1}{24 \pi }-\frac{\log 8}{24 \pi
   ^2}\right)
+\frac{1}{\ell^2}\left(\frac{1}{2 \pi ^2}-\frac{1}{8 \pi }\right)\\
&&+\left(\frac{\log \ell}{32 \pi }+\frac{\log 8\log \ell}{16 \pi ^2} -\frac{\log^28}{48 \pi ^2}-\frac{\log 8}{64 \pi }+\frac{5 \log 8}{64 \pi ^2}-\frac{9}{128
   \pi }+\frac{C}{16 \pi ^2}+\frac{1}{16 \pi ^2}+\frac{1}{64}\right)\nn
\eeqa
Where $C\simeq 0.916$ is Catalan's constant.
We see that indeed all divergent terms cancel and only the
terms with Catalan's constant survive leading to $f_{\rm 2-loop}= -C+\O(\ell^2)$
in compleat agreement with \cite{bkk}!
Note that only $2$ out of $44$ terms survive when we sum all up! This huge cancelation entangles nontrivially all the six contributions of a very different nature.
In \eq{eq:fres} we expanded $f_{\rm 2-loop}(\ell)$ further in $\ell$.

\section{Leading logarithms}
As one can see the point $\ell=0$ is a singular point of the function $f_{\rm 1-loop}$ \eq{f1loop}. The singular part is
\beq
f_{\rm 1-loop}(\ell)=-\frac{\ell^2\log\ell^2}{\sqrt{\ell^2+1}}\;.
\eeq
It contains $\log \ell$ singularity. At two loops as one can see from \eq{eq:fres} there is also $\log^2\ell$ singularity.
In this section we are aiming to understand how these singularities appear in our calculation.
The central object in our calculation is $\F(\lambda,a)$.
One can see from \eq{eq:tFof1} that with 1-loop precision,
up to regular at $a=1$ terms
\beq\la{eq:Fa}
 \F\sim -\frac{(a^2-1)}{2\sqrt\lambda}\log(a-1)+\O\(\frac{\log(a-1)}{\lambda}\)\;.
\eeq
2-loops correction in $\F$ also contains only $\log(a-1)$ to the first power as one can see from expansion in Appendix B.
This observation allows us to assume that n-loop correction will contain $\log^{n-1}(a-1)$ at
most.
Let us use this assumption about $\F$
to compute the $\log \ell$ terms to the maximal power
at each order in $1/\sqrt\lambda$.
We can use \eq{eq:al} and drop terms in r.h.s. suppressed
by $1/\lambda$, since they cannot contain $\log$ terms to the maximal power.
Concerning the leading $\log$  terms the equation
\beq
a=\sqrt{1+\ell^2+4\F(a)}\la{eq:ep}
\eeq
is exact.
For $\F$ it is enough to take 1-loop expression \eq{eq:Fa}
as far as the leading logarithms are considered. We will get
some simple quadratic equation on $a$ which leads to
\beq
a_{LL}=\sqrt{1+\frac{\ell^2}{1+2\frac{\log(a-1)}{\sqrt\lambda}}}\;.
\eeq
Using that $f=a_{LL}-\ell$ and expanding the above equation one finds
\beq\la{eq:LL}
f_{LL}=\sum_{n=0}^\infty\sum_{m=1}^\infty
\frac{k_{nm}\,\ell^{2m}\log^n\ell}{\lambda^{n/2}}\;\;,\;\;k_{nm}=(-1)^{n+m+1}
\frac{4^n(2m-3)!!\,(m+n-1)!}{2^{m}\, m!\, n!\,(m-1)!}\;.
\eeq
In particular $k_{n1}=(-1)^n4^n/2$ in agreement with \cite{am2}. The terms with $m>1$ could not be captured by the $O(6)$ sigma model. However they could correspond to a marginal
operators with many derivatives which should be added to the effective $O(6)$ sigma model action
considered by \cite{am2}.

\section{Conclusions}
In this paper we consider the $sl(2)$ sector of the AdS/CFT correspondence.
    We calculate the energy of the string rotating in $AdS_3\times S^1$
    with angular momenta $S$ and $J$
    correspondingly. In the limit $S,J\to \infty$ with
    $\ell=\frac{J\pi}{\sqrt\lambda\log S}$ fixed we compute the
    2-loop correction to its energy.

    From the gauge side of the duality this corresponds to operators of
    the form ${\rm Tr}\(D^S\Phi^J\)$ with twist $J$.
In this limit the anomalous dimensions of the operators
    scale like $J$ and one defines
    the generalized scaling function
    $f(\lambda,\ell)=\gamma(\lambda)\ell/J$. The strong coupling expansion
    of the generalized scaling function is organized in the negative
    half-integer powers of $\lambda$
    \beq
    f(\lambda,\ell)= f_{cl}(\ell)+\lambda^{-1/2}f_{\rm
    1-loop}{(1)}(\ell)+\lambda^{-1}f_{\rm 2-loop}(\ell)+\dots\;,
    \eeq
    where the first term is the classical energy-density of the string.
    The second term was computed
    in \cite{ftt,krj}. The last term is computed  in this paper as a
    function of $\ell$. Its small $\ell$  expansion reads
    \beqa\la{eq:fres}
    &&f_{\rm 2-loop}=-C+\ell^2\(8\log^2 \ell-6\log
    \ell-\frac{\log8}{2}+\frac{11}{4}\)\\ \nn
    &&+\ell^4\(-6\log^2 \ell-\frac{7\log \ell}{6}+\log 8\log
    \ell-\frac{\log^2 8}{8}+\frac{11\log
    8}{24}-\frac{233}{576}+\frac{3C}{32}\)\\ \nn
    &&+\ell^6\(6\log^2 \ell-\frac{26\log \ell}{15}-\frac{3\log 8\log
    \ell}{2}+\frac{3\log^2 8}{16}-\frac{17\log
    8}{30}+\frac{12779}{14400}-\frac{3C}{32}\)+\dots
    \eeqa
    The leading term agrees with \cite{bkk}. Also the $\ell^2\log^2\ell$
    and $\ell^2\log\ell$
    terms agree with \cite{am2} and \cite{rt2}.
    However the $\ell^2$ coefficient does not match earlier
    results of \cite{rt2}.
    It is important to understand this mismatch and to reproduce the
    higher terms in $\ell^2$  directly from the string sigma model Feynman
    diagrams.
    That will provide very a nontrivial test of the two-loop coefficient in
    the dressing phase and integrability of the $AdS_5\times S^5$ super-string sigma model.

    In this paper we also compute at each order in $1/\sqrt\lambda$ all
    the terms containing $\log\l$ to the maximal power \eq{eq:LL}
    \beqa\la{eq:llogs}
    f(\lambda,\ell)&\sim&\frac{\log
    \ell}{\lambda^{1/2}}\(-2\ell^2+\ell^4-3/4\ell^6+\dots\)\\
    \nn&+&\frac{\log^2 \ell}{\lambda}\(8\ell^2-6\ell^4+6\ell^6+\dots\)\\
    \nn&+&\frac{\log^3
    \ell}{\lambda^{3/2}}\(-32\ell^2+32\ell^4-40\ell^6+\dots\)+\dots
    \eeqa
    The $\ell^2$ terms reproduce  earlier predictions by Alday and Maldacena
    \cite{am2}.
    We have, however, a disagreement with \cite{rt2} for what concerns the $1/\lambda$ terms.

    We show that these logarithmic terms \eq{eq:llogs} are only probing
    the Hernandez-Lopez dressing phase and are not sensitive to the higher  terms in the expansion in $1/\sqrt\lambda$
    of the dressing phase. We also argue that the sub-leading logarithms could be
    computed using our method. They should be sensitive only to first few terms
    in the strong coupling expansion of the dressing phase.

    As future work, it could be interesting to compute the large $\ell$
    expansion of the scaling function. The calculation should simplify and several worldsheet loops
    could be doable. It would also be interesting to compute all
    $\log\ell$ terms in the sub-leading
    power at each order of perturbation theory and possibly check our
    results numerically.\\
    \textbf{Note added.\;\;\;}Interesting papers \cite{add0,add1,add2}
    appeared while this paper was in preparation during the last two days.
    Some of results
    seems to be similar. All these papers are based on a different approach.

\section*{Acknowledgments }
I would like to thank A.~Belitsky, A.~Jarosz, V.~Kazakov, I.~Kostov, C.~Kristjansen,
R.~Roiban, D.~Serban,
 A.~Tseytlin, A.~Vainshtein, P.~Vieira, D.~Volin and K.~Zarembo
    for many
useful discussions. The work was partially supported by RSGSS-1124.2003.2, by RFFI project
grant 06-02-16786 and ANR grant INT-AdS/CFT (contract ANR36ADSCSTZ).
I would like to thank Niels Bohr Institute and Galileo Galilei Institute where
parts of the work were done for hospitality.



\section*{Appendix A: Calculation of $c(x)$}
In this Appendix we will calculate the function $c(x)$ defined
in \eq{eq:defc} as
\beq
c^2(x)=
-16\pi\frac{g}{J}\int_a^b\frac{\varrho(y)}{ x-y}\(1-\frac{1}{y^2}\)dy\;,
\eeq
due to the suppression by $1/J$ the only chance to get
a finite result is to assume that the density for $1\ll y\ll b$ goes to a constant.
Then from large $y$'s we will get a big contribution of order $\log b\sim \log S \sim J$.
We see that to compute $c^2$ we only need some information about $\varrho(y)$
when $y$ is large. In particular for $x\sim 1$ we simply have
\beq
c^2(x)\simeq 
16\pi\frac{g}{J}\int_a^b\frac{\varrho(y)}{y}dy\;.
\eeq

To find behavior of $\varrho(x)$ for large $x$ we can still use
\eq{eq:quad}.
For $1\ll x\ll b \sim
\frac{S}{\sqrt\lambda}$ it reads
\beq
\la{ap:cc}-\frac{c^2(x)}{4}=\G^2(x)+\O(1/x)\;.
\eeq
From \eq{eq:rhoexp} we see that for large $y$ the density behaves as a constant $\varrho(y)\simeq \beta$.
Let us try to plug this into \eq{ap:cc}. What we will get is
\beq
-\frac{c^2(x)}{4}=-\frac{\pi\beta}{\ell\log S}\log (S/x)=-\frac{\pi\beta}{\ell}\(1-\frac{\log x}{\log S}\)\;.
\eeq
Whereas in the r.h.s. of \eq{ap:cc} we get $\G^2\simeq (\pi i\varrho)^2\simeq -\pi^2\beta^2$
and we see that \eq{ap:cc} cannot be satisfied  at large $x$ when $\log x\sim \log S$.
This simply means that $\varrho(x)$ is not a constant but it
could also contains terms $\frac{\log^n x}{\log^n S}$ which are not relevant when
$x$ becomes smaller. This terms are not visible in \eq{eq:rhoexp}.
In fact one can see that the only consistent with \eq{ap:cc} combination of
$\frac{\log^n x}{\log^n S}$ is
\beq
\varrho\simeq \beta_1+\beta_2\frac{\log x}{\log S}\;\;,\;\;1\ll x\ll S\;,
\eeq
integrating with this density we will get
\beq
-\frac{c^2(x)}{4}\simeq
-\frac{\pi}{\ell}\[\beta_1\(1-\frac{\log x}{\log S}\)+\frac{\beta_2}{2}\(1-\frac{\log^2 x}{\log^2S}\)\]\;.
\eeq
We have to equate this with
\beq
\G^2(x)\simeq-\pi^2\(\beta_1^2+2\beta_1\beta_2\frac{\log x}{\log S}+\beta_2^2\frac{\log^2 x}{\log^2 S}\)\;.
\eeq
Note that we get three equations on two unknowns $\beta_1$ and $\beta_2$. All of them can be resolved at the same time by setting
\beq
\beta_1=\frac{1}{2\pi \ell}\;\;,\;\;\beta_2=-\frac{1}{2\pi \ell}\;,
\eeq
so that
\beq
c^2(x)=\frac{1}{\ell^2}\frac{\log^2 (S/x)}{\log^2 (S)}
\;,
\eeq
in particular when $x\sim 1$ we get \eq{eq:cz}. 

Note that for the density we finally got
\beq
\rho\simeq\frac{J}{g}\varrho\simeq \frac{J}{2\pi g\ell}\(1-\frac{\log x}{\log S}\)=
\frac{2}{\pi}\log(S/x)
\;\;,\;\;1\ll x\ll S\;,
\eeq
which is exactly what one gets from the well-known Korchemsky's density \cite{addK}
\beq
\rho_0(u)=\frac{1}{\pi}\log\frac{1+\sqrt{1-4u^2/S^2}}{1-\sqrt{1-4u^2/S^2}}\simeq
\frac{2}{\pi}\log (S/u)\;\;,\;\;|u|\ll S,
\eeq
that can be used as an alternative derivation\footnote{We would like to thank D.Serban for pointing that out.}.

\section*{Appendix B: Expansion in $\ell$}
In Sec. \ref{sec:2loops} we expressed the 2-loop result for
the generalized scaling function $f(\lambda)$
in terms of two single integrals \eq{eq:F1res} and \eq{eq:F2res} of a rather complicated functions.
In this Appendix we give results of the expansion of these integrals
in powers of $\ell$.

Expansion of \eq{eq:F2res} reads
\beqa\la{eq:F2exp}
&&g^2\delta \F^{\rm bulk}_{\rm An}\simeq
\frac{1}{\ell^4} \left(-\frac{1}{128}+\frac{1}{8 \pi ^2}-\frac{1}{48 \pi }-\frac{\log 8}{48 \pi
   ^2}\right)
+\frac{1}{\ell^2}\left(\frac{1}{4 \pi ^2}-\frac{1}{16 \pi }\right)\\
&&+\ell^0 \left(\left(\frac{1}{64 \pi }+\frac{\log 8}{32 \pi ^2}\right) \log \ell-\frac{\log^28}{96 \pi ^2}-\frac{\log 8}{128 \pi }+\frac{5 \log 8}{128 \pi ^2}-\frac{9}{256
   \pi }+\frac{C}{32 \pi ^2}+\frac{1}{32 \pi ^2}+\frac{1}{128}\right)\nn\\
&&+\ell^2\left(\left(-\frac{3}{64 \pi ^2}+\frac{\log 8}{32 \pi ^2}\right) \log \ell-\frac{\log^28}{96 \pi ^2}+\frac{\log 8}{48 \pi ^2}-\frac{5}{768 \pi }-\frac{17}{384 \pi
   ^2}\right)\nn\\
&&+\ell^4 \left(\left(-\frac{43}{3072 \pi ^2}-\frac{3}{2048 \pi }\right) \log
   \ell-\frac{49\log 8}{18432 \pi ^2}+\frac{3\log 8}{4096 \pi } +\frac{15}{16384
   \pi }-\frac{3 C}{1024 \pi ^2}+\frac{1753}{73728 \pi ^2}\right)\nn\\
&&+\ell^6 \left(\left(\frac{113}{15360 \pi ^2}+\frac{1}{2048 \pi }\right) \log
   \ell-\frac{\log 8}{5760 \pi ^2}-\frac{\log 8}{4096 \pi } -\frac{1}{49152 \pi
   }+\frac{C}{1024 \pi ^2}-\frac{439}{76800 \pi ^2}\right)\nn\;.
\eeqa
We computed these coefficients analytically by a rather length procedure, which we do not describe here.
We checked this expansion by a numerical fit of the integral. We found that the numerical mismatch of all these coefficients is less then $10^{-45}$.

For the expansion of \eq{eq:F1res} we found
\beqa\la{eq:F1exp}
&&g^2\delta \F_{\rm HL}\simeq \frac{1}{\ell^4} \left(-\frac{1}{8 \pi ^2}+\frac{1}{48 \pi }+\frac{\log 8}{48 \pi ^2}\right)
+\frac{1}{\ell^2}\left(-\frac{1}{96}-\frac{1}{4 \pi ^2}+\frac{1}{16 \pi }\right)\\ \nn
&&+\ell^0 \left(\left(-\frac{1}{64 \pi }-\frac{5 \log 8}{32 \pi ^2}\right) \log \ell-\frac{\log^28}{192 \pi ^2}-\frac{\log 8}{128 \pi ^2}+\frac{\log 8}{128 \pi }+\frac{9}{256 \pi }-\frac{C}{16 \pi ^2}-\frac{1}{32 \pi ^2}-\frac{1}{128}\right)\\ \nn
&&+\ell^2\left(\left(\frac{7}{64 \pi ^2}-\frac{3 \log 8}{32 \pi ^2}\right) \log \ell-\frac{\log
   ^28}{192 \pi ^2}+\frac{25 \log 8}{384 \pi ^2}+\frac{5}{768 \pi }-\frac{C}{64 \pi
   ^2}+\frac{1}{12 \pi ^2}\right)\\\nn
&&+\ell^4\(\left(\frac{49}{1024 \pi ^2}+\frac{3}{2048 \pi }\right) \log \ell+\frac{493\log 8}{18432
   \pi ^2}-\frac{3\log 8}{4096 \pi }-\frac{15}{16384 \pi }+\frac{5 C}{512 \pi
   ^2}-\frac{2671}{24576 \pi ^2}\)\\ \nn
&&+\ell^6 \left(\left(-\frac{421}{15360 \pi ^2}-\frac{1}{2048 \pi }\right) \log
   \ell-\frac{1001 \log 8}{92160 \pi ^2}+\frac{\log 8}{4096 \pi } +\frac{1}{49152
   \pi }-\frac{9 C}{2048 \pi ^2}+\frac{32951}{921600 \pi ^2}\right)\;.
\eeqa
These coefficients are also checked numerically with 30 digits accuracy.

\section*{Appendix C: Main results in \textsl{Mathematica} syntaxis}
In this section we prepared the main results to be easily copied from PDF to \textsl{Mathematica}.

\begin{table}[h]
\footnotesize
\verb"tF1[x_] = -(((a^2-1)/(4 g Pi (x^2 - 1)^2)) ((2 (x^2 - 1))/(a^2 - 1) + 4 Sqrt[(a^2 - x^2)/"\\
\verb"(a^2 - 1)] Log[a^2/(a^2 - 1)] + ((2 a^2 - x^2 - 1) Log[a^4/(a^4 - 1)])/(a^2 - 1) + "\\
\verb"(((2 x^2 - a^2 (x^2 + 1))/(a^2 - 1) - 2 Sqrt[(a^2 - x^2)/(a^2 - 1)] (x^2 + 1)) "\\
\verb"Log[((x + 1) (Sqrt[a^2 - 1] x - Sqrt[a^2 - x^2]))/((x - 1) (Sqrt[a^2 - 1] x + "\\
\verb"Sqrt[a^2 - x^2]))])/ x - (I Sqrt[a^2 - x^2] (a^2 (x^2 + 1) - 2) Log[-(((Sqrt[a^2 - 1]"\\
\verb"+ I Sqrt[1 - a^2 x^2]) (I Sqrt[a^2 - x^2] + Sqrt[1 - a^2 x^2]))/((Sqrt[a^2 - 1] - "\\
\verb"I Sqrt[1 - a^2 x^2]) (Sqrt[1 - a^2 x^2] - I Sqrt[a^2 - x^2])))])/((a^2 - 1)"\\
\verb" Sqrt[1 - a^2 x^2])))"\\
\verb""\\
\verb"tF2[x_] = (2 a x (x^2 - 1) + (x^3 - 2 a^2 x + x) Log[(a - 1)/(a + 1)]"\\
\verb"+ (a^2 (x^2 + 1) - 2 x^2) Log[(a - x)/(a + x)])/(4 g Pi x (x^2 - 1)^2)"\\
\caption{
Expressions for $\t \F_{\rm HL}(x)$ and $\t \F_{\rm An}(x)$ from \protect\eq{eq:tF1}
 and \protect\eq{eq:tF2}
 }\la{tab:1}
\end{table}

\begin{table}[h]
\footnotesize
\verb"stF[x_] = (1/(g Pi (x^2 - 1)^2)) ((a - 1) (x^2 - 1) + Sqrt[(x^2 - a^2)/(a^2 x^2 - 1)] "\\
\verb"(2 - a^2 (x^2 + 1)) ArcTan[Sqrt[(x^2 - a^2)/(a^2 x^2 - 1)]] + 2 (x^2 + 1) "\\
\verb"Sqrt[((a^2 - 1) (x^2 - a^2))/x^2] ArcTan[Sqrt[x^2 - a^2]/Sqrt[a^2 x^2 - x^2]] +"\\
\verb"(1/2) (-2 a^2 + x^2 + 1) Log[(a - 1)/(a + 1)] + (1/2) (-2 a^2 + x^2 + 1)"\\
\verb" Log[a^4/(a^4 - 1)] + ((a^2 (x^2 + 1))/(2 x) - x) Log[(x + 1)/(x - 1)] +"\\
\verb"((a^2 (x^2 + 1))/(2 x) - x) Log[(x - a)/(a + x)])"\\
\caption{Expression for $\t\F(x+i0)+\t\F(x-i0)$ from \protect\eq{eq:stF}
}\la{tab:2}
\end{table}

\begin{table}[h]
{\footnotesize
\verb"Off[Series::ztest, NIntegrate::slwcon];"\\
\verb"dF2bulk[1, a0_] := -(2/g^2) NIntegrate[Re[(g stF[y] y^2)/(4 Pi (y^2 - 1)^2) /. a -> a0],"\\
\verb"{y, a0, Infinity}, WorkingPrecision -> 20, MaxRecursion -> 40]"\\
\verb""\\
\verb"c[y_]=(y^2/(l (y^2-1)^2)) (2 Sqrt[a^2-1]-Sqrt[a^2-y^2]-(a^2-1)/Sqrt[a^2-1/y^2]);"\\
\verb"dG[x_] = -((x (tF1[x] + tF2[x]))/(l Sqrt[a^2 - x^2]));"\\
\verb"dG[1] = Normal[Simplify[Series[dG[x] /. a -> Zeta[3],{x, 1, 0}] /. Zeta[3] -> a]];"\\
\verb"dF1[1, a0_] := (1/(Pi g^2)) Re[NIntegrate[g l^2 c[y] (dG[1]/y - dG[y]) (1/y) /."\\
\verb"a -> a0, {y, I, 1 - 10^(-30)}, WorkingPrecision -> 30, MaxRecursion -> 40]]"\\
\verb""\\
\verb"dF[1,l_]:=(40 l^4+48 l^2+15)/(1536 g^2 l^6)+dF1[1,Sqrt[l^2+1]]+dF2bulk[1,Sqrt[l^2+1]]"\\
\verb""\\
\verb"tF[1] = Normal[Simplify[Series[tF1[x] + tF2[x] /. a-> Zeta[3],{x,1,0}]]/.Zeta[3]->a];"\\
\verb"f2loop[l_]:=((16 Pi^2 g^2)/Sqrt[l^2+1]) ((2 D[tF[1]^2,a])/Sqrt[l^2+1]-(2 tF[1]^2)/(l^2+1)"\\ \verb"-(1/g^2) (5/(256 l^6) + 3/(64 l^4) + 1/(32 l^2)) + 2 dF[1, l]) /. a -> Sqrt[l^2 + 1]; "\\
}\la{tab:3}
\caption{Numerical evaluation of $\delta\F(a_0)$ from \protect\eq{eq:resFF}
and $f_{\rm 2-loop}(\ell)$
from \protect\eq{eq:a2}}
\end{table}



\newpage
\begin{thebibliography}{20}


\bibitem{SH38}
  M.~Staudacher,
  ``The factorized S-matrix of CFT/AdS,''
  JHEP {\bf 0505} (2005) 054
  [arXiv:hep-th/0412188].
  $\bullet$
   N.~Beisert,
  ``The su(2|2) dynamic S-matrix,''
  arXiv:hep-th/0511082.

\bibitem{SH5}
N.~Beisert and M.~Staudacher,
  ``Long-range PSU(2,2|4) Bethe ansaetze for gauge theory and strings,''
  Nucl.\ Phys.\  B {\bf 727} (2005) 1
  [arXiv:hep-th/0504190].


\bibitem{be}
B.~Eden and M.~Staudacher,
  ``Integrability and transcendentality,''
  J.\ Stat.\ Mech.\  {\bf 0611}, P014 (2006)
  [arXiv:hep-th/0603157].

\bibitem{bes}
N.~Beisert, B.~Eden and M.~Staudacher,
  ``Transcendentality and crossing,''
  J.\ Stat.\ Mech.\  {\bf 0701}, P021 (2007)
  [arXiv:hep-th/0610251].


\bibitem{AFS}
  G.~Arutyunov, S.~Frolov and M.~Staudacher,
  ``Bethe ansatz for quantum strings,''
  JHEP {\bf 0410} (2004) 016
  [arXiv:hep-th/0406256].

\bibitem{BHL}
  N.~Beisert, R.~Hernandez and E.~Lopez,
  ``A crossing-symmetric phase for AdS(5) x S**5 strings,''
  JHEP {\bf 0611} (2006) 070
  [arXiv:hep-th/0609044].

\bibitem{Maldacena}
  J.~M.~Maldacena,
  ``The large N limit of superconformal field theories and supergravity,''
  Adv.\ Theor.\ Math.\ Phys.\  {\bf 2}, 231 (1998)
  [Int.\ J.\ Theor.\ Phys.\  {\bf 38}, 1113 (1999)]
  [arXiv:hep-th/9711200].
$\bullet$
  E.~Witten,
  ``Anti-de Sitter space and holography,''
  Adv.\ Theor.\ Math.\ Phys.\  {\bf 2}, 253 (1998)
  [arXiv:hep-th/9802150].



\bibitem{bgk}
  A.~V.~Belitsky, A.~S.~Gorsky and G.~P.~Korchemsky,
  ``Logarithmic scaling in gauge / string correspondence,''
  Nucl.\ Phys.\ B {\bf 748}, 24 (2006)
  [hep-th/0601112].

\bibitem{ftt}
S.~Frolov, A.~Tirziu and A.~A.~Tseytlin,
  ``Logarithmic corrections to higher twist scaling at strong coupling from
  AdS/CFT,''
  Nucl.\ Phys.\  B {\bf 766}, 232 (2007)
  [arXiv:hep-th/0611269].

\bibitem{gkp}
  S.~S.~Gubser, I.~R.~Klebanov and A.~M.~Polyakov,
  ``A semi-classical limit of the gauge/string correspondence,''
  Nucl.\ Phys.\ B {\bf 636}, 99 (2002)
  [hep-th/0204051].


\bibitem{ft1}
  S.~Frolov and A.~A.~Tseytlin,
  ``Semiclassical quantization of rotating superstring in AdS(5) x S(5),''
  JHEP {\bf 0206}, 007 (2002)
  [hep-th/0204226].


  \bibitem{bfst}
  N.~Beisert, S.~Frolov, M.~Staudacher and A.~A.~Tseytlin,
  ``Precision spectroscopy of AdS/CFT,''
  JHEP {\bf 0310}, 037 (2003)
  [hep-th/0308117].


\bibitem{kot}
A.~V.~Kotikov, L.~N.~Lipatov, A.~I.~Onishchenko and V.~N.~Velizhanin,
  ``Three-loop universal anomalous dimension of the Wilson operators in N =  4
  SUSY Yang-Mills model,''
  Phys.\ Lett.\  B {\bf 595}, 521 (2004)
  [Erratum-ibid.\  B {\bf 632}, 754 (2006)]
  [arXiv:hep-th/0404092].



\bibitem{bern}
Z.~Bern, M.~Czakon, L.~J.~Dixon, D.~A.~Kosower and V.~A.~Smirnov,
  ``The Four-Loop Planar Amplitude and Cusp Anomalous Dimension in Maximally
  Supersymmetric Yang-Mills Theory,''
  Phys.\ Rev.\  D {\bf 75}, 085010 (2007)
  [arXiv:hep-th/0610248].
$\bullet$ F.~Cachazo, M.~Spradlin and A.~Volovich,
  ``Four-Loop Cusp Anomalous Dimension From Obstructions,''
  Phys.\ Rev.\  D {\bf 75}, 105011 (2007)
  [arXiv:hep-th/0612309].


\bibitem{ben}
M.~K.~Benna, S.~Benvenuti, I.~R.~Klebanov and A.~Scardicchio,
  ``A test of the AdS/CFT correspondence using high-spin operators,''
  Phys.\ Rev.\ Lett.\  {\bf 98}, 131603 (2007)
  [arXiv:hep-th/0611135].


\bibitem{ald}
A.~V.~Kotikov and L.~N.~Lipatov,
  ``On the highest transcendentality in N = 4 SUSY,''
  Nucl.\ Phys.\  B {\bf 769}, 217 (2007)
  [arXiv:hep-th/0611204].
$\bullet$ L.~F.~Alday, G.~Arutyunov, M.~K.~Benna, B.~Eden and I.~R.~Klebanov,
  ``On the strong coupling scaling dimension of high spin operators,''
  JHEP {\bf 0704}, 082 (2007)
  [arXiv:hep-th/0702028].
$\bullet$ I.~Kostov, D.~Serban and D.~Volin,
  ``Strong coupling limit of Bethe ansatz equations,''
  Nucl.\ Phys.\  B {\bf 789}, 413 (2008)
  [arXiv:hep-th/0703031].
$\bullet$ M.~Beccaria, G.~F.~De Angelis and V.~Forini,
  ``The scaling function at strong coupling from the quantum string Bethe
  equations,''
  JHEP {\bf 0704}, 066 (2007)
  [arXiv:hep-th/0703131].
$\bullet$ I.~Kostov, D.~Serban and D.~Volin,
  ``Functional BES equation,''
  arXiv:0801.2542 [hep-th].
$\bullet$
  D.~Fioravanti, P.~Grinza and M.~Rossi,
  ``Strong coupling for planar ${\cal N}=4$ SYM theory: an all-order result,''
  arXiv:0804.2893 [hep-th].

\bibitem{krj}
P.~Y.~Casteill and C.~Kristjansen,
  ``The Strong Coupling Limit of the Scaling Function from the Quantum   String
  Bethe Ansatz,''
  Nucl.\ Phys.\  B {\bf 785}, 1 (2007)
  [arXiv:0705.0890].
$\bullet$
A.~V.~Belitsky,
  ``Strong coupling expansion of Baxter equation in N=4 SYM,''
  Phys.\ Lett.\  B {\bf 659}, 732 (2008)
  [arXiv:0710.2294 [hep-th]].

\bibitem{rtt}
R.~Roiban, A.~Tirziu and A.~A.~Tseytlin,
``Two-loop world-sheet corrections in $AdS_5 \times  S^5$ superstring,''
  JHEP {\bf 0707}, 056 (2007)
  [arXiv:0704.3638].




\bibitem{am2}
 L.~F.~Alday and J.~M.~Maldacena,
  ``Comments on operators with large spin,''
  JHEP {\bf 0711}, 019 (2007)
  [arXiv:0708.0672 [hep-th]].



\bibitem{bkk}
  B.~Basso, G.~P.~Korchemsky and J.~Kotanski,
  ``Cusp anomalous dimension in maximally supersymmetric
  Yang-Mills theory at strong coupling,''
  arXiv:0708.3933.



\bibitem{rt}
	R.~Roiban and A.~A.~Tseytlin,
  ``Strong-coupling expansion of cusp anomaly from
  quantum superstring,'' arXiv:0709.0681.

\bibitem{rt2}
  R.~Roiban and A.~A.~Tseytlin,
  ``Spinning superstrings at two loops: strong-coupling corrections to
  dimensions of large-twist SYM operators,''
  Phys.\ Rev.\  D {\bf 77} (2008) 066006
  [arXiv:0712.2479 [hep-th]].

\bibitem{SH0}
 L.~Freyhult, A.~Rej and M.~Staudacher,
  ``A Generalized Scaling Function for AdS/CFT,''
  arXiv:0712.2743 [hep-th].

\bibitem{KMMZ}
B.~Sutherland, ``Low-Lying Eigenstates of the One-Dimensional Heisenberg Ferromagnet
for any Magnetization and Momentum,'' Phys. Rev. Lett. 74, 816 (1995)
$\bullet    $
  V.~A.~Kazakov, A.~Marshakov, J.~A.~Minahan and K.~Zarembo,
  ``Classical / quantum integrability in AdS/CFT,''
  JHEP {\bf 0405} (2004) 024
  [arXiv:hep-th/0402207].
$\bullet    $
  V.~A.~Kazakov and K.~Zarembo,
  ``Classical / quantum integrability in non-compact sector of AdS/CFT,''
  JHEP {\bf 0410} (2004) 060
  [arXiv:hep-th/0410105].

\bibitem{GK}
N.~Gromov and V.~Kazakov,
  ``Double scaling and finite size corrections in sl(2) spin chain,''
  Nucl.\ Phys.\  B {\bf 736}, 199 (2006)
  [arXiv:hep-th/0510194].

\bibitem{HernandezLopez}
  N.~Beisert and A.~A.~Tseytlin,
  ``On quantum corrections to spinning strings and Bethe equations,''
  Phys.\ Lett.\  B {\bf 629}, 102 (2005)
  [arXiv:hep-th/0509084].
$\bullet    $
  R.~Hernandez and E.~Lopez,
  ``Quantum corrections to the string Bethe ansatz,''
  JHEP {\bf 0607} (2006) 004
  [arXiv:hep-th/0603204].

\bibitem{GV2}
  N.~Gromov and P.~Vieira,
  ``Constructing the AdS/CFT dressing factor,''
  Nucl.\ Phys.\  B {\bf 790}, 72 (2008)
  [arXiv:hep-th/0703266].

\bibitem{anom}
  N.~Beisert, V.~A.~Kazakov, K.~Sakai and K.~Zarembo,
  ``Complete spectrum of long operators in N = 4 SYM at one loop,''
  JHEP {\bf 0507} (2005) 030
  [arXiv:hep-th/0503200].
$\bullet$
  N.~Beisert, A.~A.~Tseytlin and K.~Zarembo,
  ``Matching quantum strings to quantum spins: One-loop vs. finite-size
  corrections,''
  Nucl.\ Phys.\  B {\bf 715} (2005) 190
  [arXiv:hep-th/0502173].


\bibitem{addK}
  G.~P.~Korchemsky,
  Nucl.\ Phys.\  B {\bf 462} (1996) 333
  [arXiv:hep-th/9508025].

\bibitem{add0}
  B.~Basso and G.~P.~Korchemsky,
  ``Embedding nonlinear O(6) sigma model into N=4 super-Yang-Mills theory,''
  arXiv:0805.4194 [hep-th].

\bibitem{add1}
  D.~Fioravanti, P.~Grinza and M.~Rossi,
  ``The generalised scaling function: a note,''
  arXiv:0805.4407 [hep-th].

\bibitem{add2}
  F.~Buccheri and D.~Fioravanti,
  ``The integrable O(6) model and the correspondence: checks and predictions,''
  arXiv:0805.4410 [hep-th].
\end{thebibliography}
\end{document}